\documentclass[twoside]{entcs} \usepackage{entcsmacro}

\usepackage{graphicx}

\usepackage{amssymb}
\usepackage{amscd}
\usepackage{amsmath}
\usepackage{stmaryrd}
\usepackage{pifont}	
\usepackage{array}
\usepackage{listings}
\usepackage[T1]{fontenc}
\usepackage{times}
\usepackage[english]{babel}

\newcommand{\hamin}{\mathbin{\ominus}}
\newcommand{\entail}{\mathrel{\mbox{\rm :-}}}

\newcommand{\calC}{\mathcal{C}}

\newcommand{\calI}{\mathcal{I}}

 \newcommand{\calP}{\mathcal{P}}

\newcommand{\calU}{\mathcal{U}}

\newcommand{\sem}[1]{[\![#1]\!]}                
\newcommand{\RT}{\mbox{\textit{RT}}}            
\newcommand{\RTzero}{\mbox{\textit{RT$_0$}}}    
\newcommand{\RTmin}{\mbox{\textit{RT}$_{\mkern-3mu\ominus}$}}   

\def\lastname{Czenko, Tran, Doumen, Etalle, Hartel, den Hartog}
\begin{document}
\begin{frontmatter}
  \title{Nonmonotonic Trust Management for P2P Applications}
  \author{Marcin Czenko\thanksref{ishare}\thanksref{email_mczenko}},
  \author{Ha Tran\thanksref{email_htran}},
  \author{Jeroen Doumen\thanksref{ishare}\thanksref{email_jdoumen}},
  \author{Sandro Etalle\thanksref{ishare}\thanksref{inspired}\thanksref{email_setalle}},
  \author{Pieter Hartel\thanksref{email_phartel}},
  \author{Jerry den Hartog\thanksref{inspired}\thanksref{email_jdhartog}},
  \address{Department of Computer Science\\
  University of Twente\\
  Enschede, The Netherlands}
  \thanks[ishare]{This work was partially supported by the BSIK Freeband project I-SHARE.}
  \thanks[inspired]{This work was partially supported by the the EU project INSPIRED  IST-1-507894-IP.}
  \thanks[email_mczenko]{Email:
  \href{mailto:Marcin.Czenko@utwente.nl} {\texttt{\normalshape
        Marcin.Czenko@utwente.nl}}}
  \thanks[email_htran]{Email:
  \href{mailto:Ha.Tran@utwente.nl} {\texttt{\normalshape
        Ha.Tran@utwente.nl}}}
  \thanks[email_jdoumen]{Email:
  \href{mailto:Jeroen.Doumen@utwente.nl} {\texttt{\normalshape
        Jeroen.Doumen@utwente.nl}}}
  \thanks[email_setalle]{Email:
  \href{mailto:Sandro.Etalle@utwente.nl} {\texttt{\normalshape
        Sandro.Etalle@utwente.nl}}}
  \thanks[email_phartel]{Email:
  \href{mailto:Pieter.Hartel@utwente.nl} {\texttt{\normalshape
        Pieter.Hartel@utwente.nl}}}
  \thanks[email_jdhartog]{Email:
  \href{mailto:Jerry.denHartog@utwente.nl} {\texttt{\normalshape
        Jerry.denHartog@utwente.nl}}}
\begin{abstract}
  Community decisions about access control in virtual communities are
  non-monotonic in nature.  This means that they cannot be expressed in
  current, monotonic trust management languages such as the family of Role
  Based Trust Management languages (RT).  To solve this problem we propose
  $\RTmin$, which adds a restricted form of negation to the standard RT
  language, thus admitting a controlled form of non-monotonicity.  The
  semantics of $\RTmin$ is discussed and presented in terms of the well-founded
  semantics for Logic Programs.  Finally we discuss how chain discovery can be
  accomplished for $\RTmin$.
\end{abstract}
\end{frontmatter}

\section{Introduction}
\label{secIntro}

Languages from the family of Role Based Trust Management Framework (RT), like
most Trust Management (TM) languages are monotonic: adding a credential to the
system can only result in the granting of additional privileges. Usually, this
property is desirable in policy languages~\cite{SWY+02}. However, banishing
negation from an access control language is not a realistic option. In fact, as
stated by Li et al.~\cite{LFG99} ``many security policies are non-monotonic, or
more easily specified as non-monotonic ones''; similar views are expressed by
Barker and Stuckey~\cite{BS03} and by Wang et al.~\cite{WWJ04} in the context
of logic-based access control. This is also true for complex distributed
systems such as virtual communities. In particular, as we will show, modelling
access control decisions by a community, as opposed to access control decisions
by an individual member, cannot be made without at least a form of negation,
which we call negation-in-context.  As pointed out by Dung and
Thang~\cite{DT04} a TM system should be monotonic with respect to the
credential submitted by the client but could be non-monotonic with respect to
the site's local information about the client. Our extension allows a TM system
to be non-monotonic not only in a local setting, but also when the context for
negation can be provided.

\paragraph{Contributions}
We present a significant enhancement to the power of the RT family of trust
management languages by proposing $\RTmin$, an extension of $\RTzero$. More
specifically we:
\begin{itemize}
\item add a single new statement type adding negation-in-context to standard RT;
\item present and discuss the declarative semantics of $\RTmin$;
\item show that the extension is essential to specify access control policies 
      for virtual communities.
\item describe a chain discovery algorithm for $\RTmin$. 
\end{itemize}
Currently, we are using $\RTmin$ to specify and implement virtual community packages
in the context of the Freeband project \mbox{I-SHARE}.
\noindent 
In the next section we discuss how access control policies in virtual
communities motivate us to add negation-in-context to RT. In Section
\ref{secRTmin} the syntax and informal semantics of $\RTmin$ is introduced. The
formal semantics of $\RTmin$ is presented in Section \ref{secSem}. We present
related work in Section \ref{secRW} and conclusions and future work in Section
\ref{secConcl}.

\section{Virtual Communities}
\label{vc1}

Virtual communities are groups of individuals with a shared interest,
relationship or fantasy \cite{LVL02}. The majority of current virtual
communities is interested in sharing audio/video content using P2P systems
\cite{PGE+05}. Taking into account the distributed nature of virtual
communities, special mechanisms for access control must be provided to ensure
secure operations at both intra- and inter-community levels. As it is often
impossible to identify strangers \cite{PWF+02}, trust must be established
between community members and entities from outside the community prior to
allowing a specific access. We adopt the solution of SPKI/SDSI \cite{CEE+01},
where cryptographic keys are identified instead of entities.  This assumes that
each entity is the sole holder of a particular key.  As we do not want to
impose a heavy PKI, the initial trust in a new key will be low, but this trust
will increase over time (with good behaviour).

\par\noindent As an example imagine that Alice ($A$), Bob ($B$), and Carol
($C$) decide to form a virtual community (or just a community for short). At
the beginning they are the only members of the community, but they welcome
others to join. We represent a community by a list with an entry for each
member. Each entry names the community member and the members it knows about.
This knowledge results from previous interactions with the community members.
In this paper, however, when we say that one knows another community member we
we mean that one is capable of finding this member later if necessary.  Thus,
the ``knows'' relation is not necessarily commutative, since one entity can
decide to keep track of the other, but not vice versa.  For example the
following list represents the community of Alice, Bob, and Carol:
\[ A [B,C]\; B [A,C]\; C [A,B] \]
\noindent
In this community all members know each other, which means that each member can
locate any other member when needed. As the community grows it becomes harder
and harder for each member to have complete information about all other
members. Yet the community would like to protect its integrity.  Rather than to
require involvement of all members in decision making, a more practical and
scalable approach is to allow decisions about membership to be taken by a group
of coordinators selected from the community members. This group of coordinators
itself forms a (sub)community. To find all the coordinators we require that the
directed graph formed by the "knows" relation is strongly connected. This means
that each coordinator has a relationship with \textit{at least} one other
coordinator in such a way that all coordinators can be reached. For example in
the list below $A$ knows $B$, $B$ knows $C$ and $C$ knows $B$ and $A$: 
\[ A [B]\; B [C]\; C [B,A] \] 
\noindent
To become a member of a community or to become a new coordinator \textit{all}
the existing coordinators of a given community must approve.
\noindent
Trust management languages based on logic programming semantics do not support
queries of this kind directly. If one wants to know ``if all coordinators
approve entity $A$'' without explicitly enumerating these coordinators, one
must check if the \textit{negation} of this statement - ``is there any
coordinator that does not approve entity $A$'' - holds. If not, one can
conclude that all coordinators approve entity $A$. Existing trust management
languages \cite{LMW02} are strictly monotonic, thus do not allow for negation.
For this reason they are not sufficiently expressive to efficiently model
complex collaborations that commonly appear in virtual communities.
\par\noindent Before we can elaborate on this using the example just presented,
we need to review the definition of $\RTzero$, and then present our 
extension~$\RTmin$.

\section{\texorpdfstring{$\RTmin$}{RT-}}      
\label{secRTmin}

\subsection{The \texorpdfstring{$\RTzero$}{RT0} language}

$\RTzero$ contains two basic elements: \emph{entities} and \emph{role names}.
Entities represent uniquely identified principals, individuals, processes,
public keys, etc. Entities are denoted by names starting with an uppercase
letter, for example: $A$, $B$, $D$, and $Alice$. A role name begins with a
lower case letter. In $\RTzero$, roles are denoted by the entity name followed
by the role name, separated by a dot. For instance $A.r$ and
$\text{Company}.\text{testers}$ are roles. To define role membership, $\RTzero$
provides four kinds of policy statements:
\\
$\bullet$ $A.r$ $\longleftarrow$ $D$ \emph{(Simple Membership)}.  
Entity $D$ is a member of the role $A.r$.  
\\ 
$\bullet$ $A.r$ $\longleftarrow B.r_1$ \emph{(Simple Inclusion)}.
Every member of $B.r_1$ is also a member of $A.r$. This represents delegation
from entity $A$ to entity $B$. 
\\
$\bullet$ $A.r$ $\longleftarrow$ $A.r_1.r_2$ \emph{(Linking Inclusion)}.
For every entity $X$ who is a member of $A.r_1$, every member of $X.r_2$ is
also a member of $A.r$. This statement represents a delegation from entity A to
all the members of the role $A.r_1$.  The right-hand side $A.r_1.r_2$ is called
a \emph{linked role}.
\\
$\bullet$ $A.r$ $\longleftarrow$ $B_1.r_1$ $\cap$ $B_2.r_2$ 
\emph{(Intersection Inclusion)}.
Every entity which is a member of both $B.r_1$ and $B.r_2$ is a member of
$A.r$. This statement represents partial delegation from the entity A to $B_1$
and to $B_2$. The right-hand side $B_1.r_1$ $\cap$ $B_2.r_2$ is called an
\emph{intersection} role.  In a policy statement $A.r$ $\longleftarrow$ $e$ we
call $A.r$ the head and $e$ the body. The set of policy statements having the
same head $A.r$ is called the \emph{definition} of $A.r$.

\subsection{Extending \texorpdfstring{$\RTzero$}{RT0} with negation}

$\RTzero$ and other languages from the RT framework do not support negation.
As argued in Section~\ref{vc1}, this limits expressiveness. Let us first see an
example of negation to enforce the following separation of concerns policy:
``developers cannot be testers of their own code''.  We would like to express
in $\RT$ something similar to the LP clause:
\begin{equation*}
\begin{split}
 \text{verifycode}(?A) \entail \text{tester}(?A),\mathit{not}\text{ developer}(?A).
\end{split}
\end{equation*}
where $?A$ denotes a logical variable. 
This clause states that $A$ can verify the code if $A$ is a tester and $A$ is
not the developer responsible for the code. $RT^{DT}$ - another member of the
$RT$ framework~\cite{LMW02} - supports thresholds and delegation of role
activations; to some extent, $RT^{DT}$ allows to model separation of concerns
without using negation. However, this comes at the cost of having to define
manifold roles (cumbersome to work with, in practice). In any case, the
examples we present in the sequel cannot be modelled in $RT^{DT}$.
\noindent
We define a new type of statement based on $\RTzero$ and a new role-exclusion
operator $\ominus$: 
\\
$\bullet$ $A.r\longleftarrow B_1.r_1 \ominus B_2.r_2$ \emph{(Exclusion)}
All members of $B_1.r_1$ which are not members of $B_2.r_2$ are added to $A.r$.

\noindent \newline \textbf{Example}
Using the $\ominus$ operator we can solve the separation of concerns
problem as follows:
\begin{equation}
\label{verifyEx}
\begin{split}
\text{Company}.\text{verifycode}\longleftarrow\text{Company}.\text{tester}\;\ominus\;
\text{Company}.\text{developer}.
\end{split}
\end{equation}
Suppose that both \textit{Alice} and \textit{Bob} are testers but Alice is also a developer of the code:
\begin{equation*}
\label{agreeToAddEx}
\begin{split}
\text{Company}.\text{tester}\longleftarrow\text{Alice}&\;\;\;\;
\text{Company}.\text{tester}\longleftarrow\text{Bob}\\
\text{Company}.\text{developer}&\longleftarrow\text{Alice}\\
\end{split}
\end{equation*}
We see that credential \ref{verifyEx} does not make Alice be a member of the\\
$\text{Company}.\text{verifycode}$ role. Thus, only Bob can verify the code.

\subsection{Modelling virtual communities using \texorpdfstring{$\RTmin$}{RT-}}
\label{modelVC}

Having given a simple example and its representation in RT$_{\ominus}$, we now
return to the more complex scenario of community decision making from
Section~\ref{vc1}.  
\par\noindent Recall that we have a community of coordinators - \textit{Alice}
($A$), \textit{Bob} ($B$), and \textit{Carol} ($C$). Assume that another entity
- say $D$ - wants to join this community and asks \textit{Alice} for approval.
\textit{Alice} can accept $D$ as a new coordinator locally, but before making
the final decision she must check if there is no objection from other
coordinators. A coordinator expresses the objection using a so called
\textit{black list}. An entity that is on the black list of one of the
coordinators will not be accepted as a new coordinator.
\begin{table*}[ht]
\footnotesize
\caption{Roles used by coordinators}
\label{roles}
\begin{center}
\begin{tabular}{|m{5.2cm}|m{6.4cm}|m{0.6cm}|} \hline
Definition (for coordinator $A$) & Description & 
\shortstack{\rule{0mm}{1mm}\\Op-\\tio-\\nal} \\ \hline\hline
\parbox[c]{5.2cm}{
$A.\text{agreeToAdd}\longleftarrow [\text{set of entities}]$
} &
A coordinator uses this role to express that she approves an entity.  The role
has a local meaning. It is not sufficient to be a member of the
\textit{agreeToAdd} role to become a coordinator.  It is necessary that no
other coordinators says that an entity is a member of her
\textit{disagreeToAdd} role. The \textit{agreeToAdd} role, through the
\textit{allCandidates} role, provides context for the $\ominus$ operator in the
definition of the the \textit{addCoord} role.
& \\ \hline
\parbox[c]{5.2cm}{
\vspace{.2ex}
$A.\text{disagreeToAdd}\longleftarrow$ \\
$\hspace*{1cm}[\text{see description in the text}]$ \vspace{.2ex}
} & 
This role is used by a coordinator as a black list.
& \\ \hline
\parbox[c]{5.2cm}{
$A.\text{coord}\longleftarrow [\text{set of entities}]$
} &
This role contains all the coordinators known
by a coordinator. 
&  \\ \hline
\parbox[c]{5.2cm}{
$A.\text{allCoord}\longleftarrow A$\\
$A.\text{allCoord}\longleftarrow A.\text{allCoord}.\text{coord}$
} &
This role allows a coordinator to iterate over all entities
connected by the \textit{coord} role. This role, if defined,
contains all the coordinators.
& \ding{51} \\ \hline
\parbox[c]{5.2cm}{
$A.\text{objectionToAdd}\longleftarrow$\\
\hspace*{1.2cm}$A.\text{allCoord}.\text{disagreeToAdd}$
} &
A coordinator can use this role to obtain all entities for
which there is any objection.
& \ding{51} \\ \hline
\parbox[c]{5.2cm}{
$A.\text{allCandidates}\longleftarrow$\\
\hspace*{1.5cm}$A.\text{allCoord}.\text{agreeToAdd}$
} &
This role, if defined, contains all the candidate coordinators
locally accepted by any of the coordinators. Used as the context
for the $\ominus$ operator in the body of the \textit{addCoord} role.
& \ding{51} \\ \hline
\parbox[c]{5.2cm}{
$A.\text{addCoord}\longleftarrow A.\text{allCandidates}\;\ominus$\\
\hspace*{1.5cm}$A.\text{objectionToAdd}$
} &
After becoming a member of this role, a candidate coordinator becomes
a new coordinator and becomes a member of the \textit{coord} role.
& \ding{51} \\ \hline
\end{tabular}
\end{center}
\end{table*}
\par\noindent
Table \ref{roles} shows the minimal definition, and the descriptions of the
roles used by coordinators.  We see from Table \ref{roles} that some roles are
mandatory while the others are not. For instance the role
\textit{disagreeToAdd} must be defined by each coordinator. On the other hand,
the roles \textit{allCoord}, \textit{allCandidates}, and \textit{addCoord} can
be defined as needed by a coordinator. Special attention must be given to the
definition of the \textit{disagreeToAdd} role.  For example, a coordinator can
use the following credential to say that she distrusts any entity she does not
accept locally:
\begin{equation*}
\begin{split}
A.\text{disagreeToAdd}\longleftarrow A.\text{allCandidates}\;\ominus\; A.\text{agreeToAdd}.\\
\end{split}
\end{equation*}
If a coordinator trusts other coordinators to select candidates she can leave
the \textit{agreeToAdd} role empty and use her \textit{disagreeToAdd} role to
block some candidates. For example, \textit{Alice} can put $E$ on her black
list to disallow $E$ to become a coordinator, and simultaneously accept all
other candidates proposed by other coordinators:
\begin{equation*}
A.\text{disagreeToAdd}\longleftarrow E.\\
\end{equation*}

\noindent
Table \ref{scenario} shows the roles and their members as seen by Alice, Bob,
and Carol. In this table, we assume that Alice agrees locally to add $D$ as a
new coordinator. Also, Bob and Carol have no objection to add $D$ as a new
coordinator, but $E$ is on Alice's black list and $F$ is on the black list of
Bob and Carol. As a consequence, only $D$ is the member of the
\textit{addCoord} role of Alice. Bob and Carol do not have to define the
\textit{allCoord}, \textit{allCandidates}\footnote{A coordinator must define
the \textit{allCandidates} role if she defines the \textit{disagreeToAdd} role
in terms of the \textit{agreeToAdd} role.}, \textit{objectionToAdd}, and
\textit{addCoord} unless they themselves add a new coordinator.

\begin{table*}[ht]
\footnotesize
\caption{Adding a new coordinator - $D$ is successful, $E$, $F$ fail ($ND$ = Not Defined)}
\label{scenario}
\begin{center}
\begin{tabular}{|c||c|c|c|c|c|c|c|} \hline
		& \rotatebox{90}{coord} & \rotatebox{90}{agreeToAdd} & \rotatebox{90}{allCoord} & \rotatebox{90}{allCandidates} & \rotatebox{90}{disagreeToAdd} & \rotatebox{90}{objectionToAdd\;\;} & \rotatebox{90}{addCoord} \\ \hline\hline
Alice (A)      &   \{B\}   &     \{D\}         &   \{A,B,C\}  &    \{D\}   &  \{E\}          &       \{E,F\}        &    \{D\}     \\ \hline
Bob (B)        &   \{C\}   &     \{\}       & $ND$ &  $ND$  &  \{F\}           &    $ND$  & $ND\addtocounter{mpfootnote}{-1}\mpfootnotemark{}$ \\ \hline
Carol (C)      &   \{B,A\} &     \{\}       & $ND$ &  $ND$  & \{F\}           &    $ND$  & $ND$ \\ \hline
\end{tabular}
\end{center}
\end{table*}

\section{Semantics}
\label{secSem}

The semantics of trust management languages is typically given by a translation
into Logic Programming (LP) \cite{LMW02}. We will follow the same route. Trust
management credentials are by definition distributed among different
principals. The use of negation creates an additional difficulty, also because
in logic programming various different semantics exist to cope with
negation.  We have chosen to use the Well-Founded (WF) semantics~\cite{GRS91}
for the reasons sketched below.  The WF semantics imposes no restrictions on
the syntax of programs, provides an \emph{unique} model for each program (as
opposed to e.g.\ the stable model semantics~\cite{GL88}) and enjoys an elegant
fixed-point construction.  

The WF semantics basically works as follows (we refer the interested reader
to~\cite{GRS91} for details): For a program, consisting of a set of rules, one
iteratively builds positive and negative facts.  Positive facts are obtained as
usual; any fact that can be derived by a rule from the already found facts is
added.  Negative facts are obtained from `unfounded sets' which contain currently
undecided facts which no rule can derived even when the elements of this set are 
set from undecided to false.  
Thus setting this unfounded set to false will not create contradictions.  As we
cannot always obtain a positive or negative version of each fact, some atoms
will remain undecided and be assigned the value `undefined', i.e.~the WF semantics 
is three valued.

 In a TM system it is impossible to avoid circular references, and we cannot
expect policies to be (locally) \emph{stratified}.  Stratification basically
means that one can restructure a logic program into separate parts in such a
way that negative references from one part refer only to previously defined
parts.  Without the possibility of local stratification we cannot refer to the
\emph{perfect model semantics} \cite{Prz88}.  For the same reason, we certainly
have to refer to a \emph{three valued semantics}: Next to the truth values
\emph{true} and \emph{false}, we have to admit the valued \emph{undefined}. In
short, this is because we cannot expect the completion of a policy to be a
consistent logic program in the sense described in \cite{She88}.  

The handling of positive circular references, as in $\{A.r \longleftarrow
B.r\;\;\; B.r \longleftarrow A.r\}$ should be done in accordance with the
semantics of $\RTzero$; we should obtain that some entities, for example $C$,
do \emph{not} belong to $A.r$. This forces us to exclude Kunen's
semantics~\cite{Kun87} (i.e.{} the semantics of logical consequences of the
completion of the program together with the weak domain closure assumptions),
and Fitting's semantics~\cite{Fit85}: in both semantics the query ``does $C$
belong to $A.r$?'' would return \emph{undefined}.  The WF semantics does return
false for this membership query.

\begin{example}
 Consider the program~$\calP$ with the following clauses:
 \begin{displaymath}
    p \entail q. 	\qquad 
    q \entail p.	\qquad 
    r \entail \neg q. 	\qquad 
    s \entail \neg t.	\qquad 
    t \entail \neg s.	\qquad 
    u \entail \neg s.
 \end{displaymath}
 In the well-founded model of $\calP$ we have that $p$ and $q$ are false, 
 $r$ is true and $s$, $t$, and $u$ are undefined. (On the other hand, all 
 predicates would be undefined in Kunen's semantics.)
\end{example}

\subsection{Translating \texorpdfstring{$\RTmin$}{RT-} to GLP} 
\label{subsecTransToLP}

We first give the translation to LP for $\RTzero$ and, using this translation,
the semantics of a set of $\RTzero$ policy statements.  Next we extend this to
a translation from $\RTmin$ to GLP and the semantics for a set of $\RTmin$
policy statements.

\noindent
The semantics of a set of $\RTzero$ policy statements is commonly defined by
translating it into a logic program~\cite{LMW02}.  Here, we depart from the
approach of Li et al.~\cite{LMW02} by referring to the role names as predicate
symbols. The statement $A.r$ $\longleftarrow$ $D$ is, for example, translated
to $r(A,D)$ in the Prolog program.  Intuitively, $r(A,D)$ means that $D$ is a
member of the role $A.r$.

\begin{definition}
\label{defTranslateRT}
 Given a set $\mathcal{P}$ of $\RTzero$ policy statements, the \emph{semantic
 program}, $\!SP(\mathcal{P})$, for $\mathcal{P}$ is the logic program defined
 as follows (recall that symbols starting with ``?'' represent logical
 variables):
 
 $\bullet$ For each $A.r \longleftarrow D \in \calP$ add to $SP(\calP)$ the 
 clause $r(A,D)$
 
 $\bullet$ For each $A.r \longleftarrow B.r_1 \in \calP$ add to $SP(\calP)$ the
 clause $r(A,?Z) \entail r_1(B,?Z)$
 
 $\bullet$ For each $A.r \longleftarrow A.r_1.r_2 \in \calP$ add to $SP(\calP)$ 
 the clause $r(A,?Z) \entail r_1(A,?Y),$ $r_2(?Y,?Z)$
 
 $\bullet$ For each $A.r \longleftarrow B_1.r_1 \cap B_2.r_2 \in \calP$ add to 
 $SP(\mathcal{P})$ the clause $r(A,?Z) \entail r_1(B_1,?Z), r_2(B_2,?Z)$ 
 
\label{defSemRT} 
  \noindent
  The \emph{semantics} of a
  role $A.r$ is a set of members $Z$ that make the predicate $r(A,Z)$ true in
  the semantic program: $\sem{A.r}_{\calP} = \{ Z \,\vert\,SP(\calP) \models r(A,Z) \}$
\end{definition}
\noindent 
We write $SP(P) \models r(A,Z)$ if $r(A,Z)$ is true in the unique
well-founded model of $P$. (For negation-free programs this model 
coincides with the least Herbrand model used for the semantics of $\RTzero$ 
by Li at al~\cite{LMW02}.)
We now extend the translation of $\RTzero$ to that of $\RTmin$ by adding the
translation of the exclusion rule.  
\begin{definition}
\label{defTranslateExtended}
  Given a set $\mathcal{P}$ of $\RTmin$ policy statements, the \emph{semantic
  program}, $SP(\mathcal{P})$, for $\mathcal{P}$ is the \emph{general} logic
  program defined as follows: 
  
  \noindent
  $\bullet$ For each $A.r \longleftarrow B.r_1 \ominus B.r_2 \in \calP$ 
  	add to $SP(\calP)$ the clause $r(A,?Z) \entail r_1(B_1,?Z), \neg r_2(B_2,?Z)$
\\
$\bullet$ All other rules are as in definition~\ref{defTranslateRT}.

\label{defSemExtended}
\noindent
  The \emph{semantics} of a
  role $A.r$ is a set of members $Z$ that make the predicate
  $r(A,Z)$ true in the semantic program:    
  	$\sem{A.r}_{\calP} = \{ Z \,\vert\, SP(\calP) \models r(A,Z) \}$ 
\end{definition}
\noindent 
Note that, unlike before, the value of the semantical program may give value
`undefined' for $r(A,Z)$.  In this case the agent $Z$ is not considered to be a
member of the role, nor of the negated role.  
\begin{example}
  Consider a system with entities $A,B,C,D$, roles $A.r, B.r$ and $C.r$ and the
  following policy rules:
  \[ A.r\longleftarrow B.r\hamin C.r \qquad C.r\longleftarrow B.r\hamin A.r \qquad B.r\longleftarrow D\]
  Here $D$ is a member of $B.r$, however, $D$ is not a member of either $A.r$
  or $C.r$. 
  Note that as a result we have that despite the presence of the rule
  $A.r \longleftarrow B.r \hamin C.r$ the role $B.r$ can have members that are
  neither in $A.r$ nor in $C.r$.  
\end{example}

\noindent
The rules for $A.r$ and $C.r$ in the example above are referred to as negative
circular dependencies; $A.r$ depends negatively on $C.r$ and $C.r$, in turn,
depends negatively on $A.r$.  The example shows that care is required when
reasoning about policies which have negative circular dependencies.

\subsection{Virtual Communities - translation to GLP}

Having introduced an example of virtual community decision making in
Section~\ref{vc1}, its formalism in Subsection~\ref{modelVC}, we now give the
GLP semantics of the example. Translating $\RTmin$ credentials to GLP is
straightforward using the rules presented in Subsection~\ref{subsecTransToLP}.
For the convenience of the reader we present a complete policy and the
corresponding GLP rules in \ref{appendix}. If one asks Alice to add $D$ to the
group of coordinators she needs to check if $D$ is a member of the
\textit{A.addCoord}.  This is equivalent to checking whether
\textit{addCoord(A,D)} holds after the translation to GLP. She does this by
checking whether $D$ is a logical consequence of the semantic program
$SP(\mathcal{P})$ by first finding the semantics of the role
\textit{A.addCoord} and checking if it contains entity $D$. The semantics of
the role \textit{A.addCoord} with respect to the program $\mathcal{P}$ is as
follows:
\begin{equation*}
  \llbracket A.\text{addCoord} \rrbracket_{\mathcal{P}} = \{ D \}.
\end{equation*}
\noindent
The semantics of the roles \textit{A.allCandidates} and
\textit{A.objectionToAdd} (these roles define the role \textit{A.addCoord}) are
shown below: 
\begin{equation*}
  \llbracket A.\text{allCandidates} \rrbracket_{\mathcal{P}} = \{ D \}\;\;\;
  \llbracket A.\text{objectionToAdd} \rrbracket_{\mathcal{P}} = \{ E,F \}.
\end{equation*}
\noindent
The semantics of a role may also be an empty set: $\llbracket
B.\text{agreeToAdd} \rrbracket_{\mathcal{P}} = \{ \}.$

\section{Credential Chain Discovery}
\label{secChain}

In this section we extend the standard chain discovery algorithm to $\RTmin$
following the construction of the well-founded semantics.  Recall that the
definition of a role $A.r$ is the set of all credentials with head $A.r$.  We
assume that $A$ stores (or at least, is able to find) the complete definition
of each of her roles $A.r$, i.e.~that the credentials involved are
issuer-traceable.
\noindent
The main difficulty in the chain discovery is to obtain that $B$ is not a
member of a linked role $A.r.r'$. For this we need to check that every
potential member $C$ of $A.r$ does not have $B$ in its role $C.r'$.  So who are
the potential members of $A.r$?  Thanks to negation in context we can provide a
reasonable overestimation of this set using chain discovery for $\RTzero$:
\begin{definition}
  \label{defcontext}
  For a policy $\calP$ the context policy $\calP+$ 
  is the policy obtained by replacing each credential of the form 
  $A.r \longleftarrow B_1.r_1 \ominus B_2.r_2 \in \calP$ 
  by 
  $A.r \longleftarrow B_1.r_1$ 
  and leaving the other credentials unchanged.  
  We call $\sem{A.r}_{\calP+}$ the \emph{context} of the role~$A.r$.
\end{definition}
\noindent The following lemma relates roles with their contexts.
\begin{lemma}
  \label{lemcontext}
  For any policy $\calP$ and role $A.r$ we have: If $SP(\calP) \models r(A,B)$
  then $SP(\calP+) \models r(A,B)$ and if $SP(\calP+) \not\models r(A,B)$ then
  $SP(\calP) \models \neg r(A,B)$. 
\end{lemma}
\noindent 
The first part of this lemma states that any role is contained in its context,
$\sem{A.r}_{\calP} \ \subseteq \ \sem{A.r}_{\calP+}$.  If
$B\not\in\sem{A.r}_{\calP}$ this means that $r(A,B)$ is undefined or false in
$SP(\calP)$.  The second part of the lemma states that if $B\not\in
\sem{A.r}_{\calP+}$ it must be the latter, $r(A,B)$ is false in $SP(\calP)$.
\noindent
In the algorithm below we build a set of credentials $\calC$ together with a
set of context membership facts $\calI+$ and a set of positive and negative
membership facts $\calI$.  

\noindent 
\textbf{Step 1}.  Initialise $\calI = \emptyset$, $\calI+ = \emptyset$ and 
	$\calC$ = the definition of role $A.r$. 

\noindent 
\textbf{Step 2}.  Discover context and credentials (classical chain discovery 
	for $\calI+$ and $\calC$). 

  \noindent We look for new credentials top down; any credential that could possibly be
  relevant for role $A.r$ is added to $\calC$.  We look for the context of
  $A.r$ bottom up; any fact that can be derived from the credentials that we
  have found is added to $\calI+$.  Repeat the following until no changes occur: For
  each credential of the following form in~$\calC$:

  [$B.r_0 \longleftarrow C$] 
  	add $r_0(B,C)$ to $\calI+$ 

  [$B.r_0 \longleftarrow C.r_1$] 
  	add the definition of $C.r_1$ to $\calC$ and 
	add $r_0(B,D)$ to $\calI+$ for all $r_1(C,D)$ in $\calI+$ 

  [$B.r_0 \longleftarrow C_1.r_1 \cap C_2.r_2$]
  	add the definitions of $C_1.r_1$ and $C_2.r_2$ to $\calC$
	add $r_0(B,D)$ to $\calI+$ whenever $r_1(C_1,D)$ and $r_2(C_2,D)$ in $\calI+$.

  [$B.r_0 \longleftarrow C.r_1.r_2$] 
	add the definition of $C.r_1$ and, 
	for each $r_1(C,D) \in \calI+$, the definition of $D.r_2$ to $\calC$. 
	Add $r_0(B,D)$ to $\calI+$ whenever for some $Y$ we have $r_1(C, Y)$ 
	and $r_2(Y, D)$ in $\calI+$.

  [$B.r_0 \longleftarrow C_1.r_1 \ominus C_2.r_2$]
  	add the definitions of $C_1.r_1$ and $C_2.r_2$ to $\calC$,
  	add $r_0(B,D)$ to $\calI+$ for every $r_1(C_1,D)$ 

\noindent 
\textbf{Step 3}.  Discover positive facts in $\calI$ (extended chain 
	discovery~1).

  \noindent We update $\calI$ similar to $\calI+$ in the previous step, only the last case ($\ominus$) changes.
  Repeat until $\calI$ does not change, for credentials in $\calC$ of the following form:

  [$B.r_0 \longleftarrow C$] 
  	add $r_0(B,C)$ to $\calI$ 

  [$B.r_0 \longleftarrow C.r_1$] 
	add $r_0(B,D)$ to $\calI$ for all $r_1(C,D)$ in $\calI$ 

  [$B.r_0 \longleftarrow C_1.r_1 \cap C_2.r_2$]
	add $r_0(B,D)$ to $\calI$ whenever $r_1(C_1,D)$ and $r_2(C_2,D)$ in $\calI$.

  [$B.r_0 \longleftarrow C.r_1.r_2$] 
	Add $r_0(B,D)$ to $\calI$ whenever for some $Y$ we have $r_1(C, Y)$ 
	and $r_2(Y, D)$ in $\calI$.

  [$B.r_0 \longleftarrow C_1.r_1 \ominus C_2.r_2$]
  	add $r_0(B,D)$ to $\calI$ whenever $r_1(C_1,D)\in \calI$ and 
	either $(\neg r_2(C_2,D)) \in \calI$ or $r_2(C_2,D)\not\in\calI+$.

\noindent 
\textbf{Step 4}. Discover negative facts in $\calI$ (extended 
	chain discovery~2).

  \noindent 
  We search for facts which are useful when negated in $\calI$: Initialise
  $\calU = \emptyset$.  We say an atom $r(X,Y)$ is \emph{not yet false (NYF)}
  if it is a member of the context and not assumed or known to be false, i.e.{}
  $r(X,Y)\in \calI+$, $r(X,Y) \not\in \calU$ and $\neg r(X,Y) \not\in \calI$.
  A fact $r_2(C_2,D)$ is useful if it is not yet false and $\neg r_2(C_2,D)$
  can be used to derive a fact, i.e.~$B.r_0 \longleftarrow C_1.r_1 \ominus
  C_2.r_2 \in \calC$ and $r_1(C_1,D) \in \calI$.  Choose one useful fact and
  add it to $\calU$.

  \noindent
  Next we try to show that facts in $\calU$ are false by showing that no rule
  can possibly derive a fact in $\calU$.  To achieve this we may need to assume
  that other facts are also false, i.e.~add them to $\calU$. \\ For each fact
  $r(B,D)$ in $\calU$ and matching rule $B.r \longleftarrow e\in\calC$ perform:

  [$B.r \longleftarrow C$] Do nothing.

  [$B.r \longleftarrow C.r_1$] 
	This rule cannot be used to derive $r(B,D)$ if $r_1(C,D)$ is false thus
	if $r_1(C,D)$ is NYF then add it to $\calU$.

  [$B.r \longleftarrow C_1.r_1 \cap C_2.r_2$] 
	If $r_1(C_1,D)$ and $r_2(C_2,D)$ are both NYF then choose one to add to
	$\calU$.

  [$B.r \longleftarrow C_1.r_1 \ominus C_2.r_2$]
	If $r_1(C_1,D)$ is NYF and $r_2(C_2,D) \not\in \calI$ then add
	$r_1(C_1,D)$ to $\calU$.

  [$B.r \longleftarrow C.r_1.r_2$]
	For all $Y$ with $r_1(C,Y)$ NYF: If $r_2(Y,D)$ is NYF choose one of
	$r_1(C,Y)$ and $r_2(Y,D)$ and add it to $\calU$. 

  $\star$ Try each possible choice in the substep above and if the resulting
  	  $\calU$ has no elements in common with $\calI$ then add $\neg\, \calU$
	  to $\calI$.

\noindent 
Repeat steps 3 and 4 until $\calI$ remains unchanged.

\noindent
(End of algorithm.)
The algorithm correctly finds the members of the role $A.r$:
$$\forall B: \  r(A,B) \in \calI  \iff  B \in \sem{A.r}_{\calP}$$
It follows the steps in the construction of the well-founded semantics 
in such a way that $\calI$ is, at each stage, a sufficiently large 
subset of the well-founded model.

\section{Implementation}
\label{secImp}

In the current prototype storage is centralised and we assume that all
credentials can be traced by the issuer. In such a case, Linear resolution with
Selection function for General logic programs (SLG) resolution of XSB prolog
can be used to compute answers to queries according to the WF model for
$\RTmin$~\cite{CSW95}.  XSB is a research-oriented, commercial-grade Logic
Programming system for Unix and Windows-based platforms. XSB provides standard
prolog functionality but also supports negations and constraints. Using SLG
resolution XSB prolog can correctly answer queries for which standard prolog
gets lost in an infinite branch of a search tree, where it may loop infinitely.
A number of interfaces to other software systems including Java and ODBC are
available.  DLV datalog~\cite{EFL+00} and the Smodels system~\cite{NS97} can
also be used to provide an initial implementation of $\RTmin$. The DLV
system~\cite{EFL+00} is a system for disjunctive logic programs. It is
distributed as a command line tool for both Windows and Linux operation
systems. DLV is capable of dealing with disjunctive logic programs without
function symbols allowing for strong negations, constraints and queries.  DLV
uses two different notions of negation: negation as failure and true (or
explicit) negation. By default, DLV handles negation as failure by constructing
the stable model semantics for the program. This standard behaviour can be
changed using a command line option and then a WF model is built instead. The
true or explicit negation expresses the facts that explicitly are known to be
false. On the contrary, negation as failure does not support explicit assertion
of falsity. Models of programs containing true negation are also called
``answer sets''.  The Smodels system~\cite{NS97} provides an implementation of
the well-founded and stable model semantics for range-restricted function-free
normal programs.  The Smodels system allows for efficient handling of
non-stratified ground programs and supports extensions including built-in
functions, cardinality, and weight constraints. The Smodels system is available
either as a C++ library that can be called from user programs or as a
stand-alone program with default front-end (\textit{lparse}).
\noindent
We implemented the program introduced in sub-section 4.3 on three systems: XSB,
Smodels and DLV. To test the performance of the program on these systems, we
use two parameters: number of coordinators (Coords) and number of iterations
(Iters). The higher the number of coordinators is, the more complex the program
is. The program is also executed repeatedly to compare performance more
correctly. Table~\ref{result} in the appendix reports the execution time of the
program measured by the CPU time obtained.    
\noindent
We cannot compare the execution time between XSB and the other two DLV and
Smodels because XSB is the goal-oriented system while DLV and Smodels build and
return the whole model for the program. Because of this XSB is faster than the
other two systems. DLV provides better execution time than Smodels, especially
when the complexity of the program increases.

\section{Related Work\label{secRW}}

So far little attention has been given to trust management in virtual
communities. Most of the existing approaches focus on reputation-based trust
models in P2P networks~\cite{ST05}. Abdul-Rahman and Hailes~\cite{AH00} propose
a trust model that is based on real world social trust characteristics. They
also find formal logic based trust management to be ill suited as a general
model of trust. To prove this claim they refer to the early work of Burrows and
Abadi~\cite{BAN90}, and Gong, Needham, and Yahalom~\cite{GNY90}, which are more
relevant to formal protocol verification than to formal reasoning on trust
management. To support their work they claim that logic based trust management
systems are not suitable to be automated - the existing literature on automated
trust negotiation (ATN) yields a contradictory statement (see Seamons et
al.~\cite{SWY+02}).  Pearlman et al.~\cite{PWF+02} present a Community
Authorisation Service - a central management unit for a community that helps to
enforce the policy of a virtual community. Such a central point of
responsibility does not fit well in the spirit of P2P networks because of their
highly distributed nature.  Pearlman et al.{} also require that there a
centralised policy exists for a virtual community. However, the policy of a
virtual community may have a distributed character and can be seen as a product
of the policies of the community members. Boella and van der Torre~\cite{BT04}
take the same direction and emphasise the distinction between authorisations
given by the Community Authorisation Service and permissions granted by
resource providers in virtual communities of agents. They regard authorisation
as a means used by community authorities to regulate the access of customers to
resources that are not under control of these authorities. According to Boella
and van der Torre, permission can be granted only by the actual resource owner.

As we conclude in Section~\ref{vc1}, virtual communities are also not supported
by the existing trust management languages, even though the general
requirements for such languages have been investigated~\cite{SWY+02}.

Herzberg et al.\ propose in \cite{HMM+00} a prolog-based trust management
language (DTPL) together with a non-monotonic version of it (TPL).  Their
approach is very different from ours in the sense that TLP allows for
\emph{negative certificates} namely ``certificates which are interpreted as
suggestions not to trust a user''. This far-reaching approach leads to a more
complex logical interpretation, which includes conflict resolution. As opposed
to this, our approach is technically simpler and enjoys a well-established
semantics.
\noindent
Jajodia et al.~\cite{JSS+01}, Wang et al.~\cite{WWJ04}, Barker and
Stuckey~\cite{BS03}, have in common that they impose a \emph{stratified} use of
negation. Because of this, they can refer to the perfect model semantics. As we
explained in Section \ref{secSem}, in the context of DTM, we cannot expect
policies to be stratified. Our approach is thus more powerful than the
approaches based on the stratifiable negation.
\noindent
Dung and Thang in \cite{DT04} propose a DTM system based on logic programming
and the \emph{stable model semantics} \cite{GL88}.

\section{Conclusions and future work}
\label{secConcl}

We present the language $\RTmin$, which adds a construct for
`negation-in-context' to the $\RTzero$ trust management system.  We argue the
necessity of such a construct and illustrate its use with scenarios from
virtual communities which cannot be expressed within the $\RT$ framework.  

We provide a semantics for $\RTmin$ by translation to general logic programs.
We show that, given the complete policy, the membership relation can be decided
by running the translation in systems such as XSB, DLV datalog and Smodels.  We
also show how, for the case that credentials are issuer traceable~\cite{LWM03},
the chain discovery algorithm for $\RTzero$ can be extended to $\RTmin$.  We
are currently employing $\RTmin$ to specify virtual community policies in the
Freeband project I-SHARE.
In the future we plan to examine the complexity of the presented chain
discovery algorithm, ad hoc methods to minimise communication overhead, and
safe methods for chain discovery in non-`issuer traces all' scenarios.  A
comparison with reputation systems will also be made.

In section~\ref{secChain} we have assumed that the credentials are issuer
traceable and that we are able to obtain all relevant credentials.  In our
scenario this is realistic; as the coordinators play a central role, they are
generally assumed to be available sufficiently often and have sufficient
resources to store their own credentials.  In general collecting all
credentials can be difficult, for example, credentials may be stored elsewhere,
entities may be unreachable or messages may be lost.  In such a situation, we
cannot safely determine that $A$ is not in $B$'s role $r$ by absence of
credentials.  Instead we could ask $B$ to explicitly state that $A$ is
\emph{not} a member of $B.r$.  This is sufficient if we know the context of a
role (and thus which negative facts we need).  More advanced mechanisms to
guarantee safety of roles and a precise definition of which policies are safe
using which mechanism is subject of further research.

\bibliographystyle{plain}
\bibliography{bibfile}

\appendix
\renewcommand\thesection{Appendix \Alph{section}}
\section{\texorpdfstring{}{Appendix A}\label{appendix}}
\renewcommand\thesection{\Alph{section}}

\begin{table}[ht]
\footnotesize
\begin{center}
\caption{Virtual Community - translation to GLP}
\begin{tabular}{|l|l|} \hline
RT$_\ominus$\, rules & GLP semantics \\ \hline
\parbox[b][9.7cm][t]{5.8cm}{
\begin{equation*}
\begin{split}
& A.\text{addCoord}\longleftarrow A.\text{allCandidates}\;\ominus\\
&\hspace{1cm} A.\text{objectionToAdd}\\
& A.\text{allCandidates}\longleftarrow\\
&\hspace{1cm} A.\text{allCoord}.\text{agreeToAdd}\\
& A.\text{objectionToAdd}\longleftarrow\\
&\hspace{1cm} A.\text{allCoord}.\text{disagreeToAdd}\\
& A.\text{disagreeToAdd}\longleftarrow A.\text{allCandidates}\;\ominus\\
&\hspace{1cm} A.\text{agreeToAdd}\\
& A.\text{allCoord}\longleftarrow A.\text{allCoord}.\text{coord}\\\\
& A.\text{allCoord}\longleftarrow A\\
& A.\text{coord}\longleftarrow B\\
& B.\text{coord}\longleftarrow C\\
& C.\text{coord}\longleftarrow B\\
& C.\text{coord}\longleftarrow A\\
& A.\text{agreeToAdd}\longleftarrow D\\
& A.\text{disagreeToAdd}\longleftarrow E\\
& B.\text{disagreeToAdd}\longleftarrow F\\
& C.\text{disagreeToAdd}\longleftarrow F\\
\end{split}
\end{equation*}
} & \parbox[b][9.7cm][t]{6.7cm}{
\begin{equation*}
\begin{split}
&\text{addCoord}(A,?Y)\text{:- allCandidates}(A,?Y),\\
&\hspace{1cm}\neg \text{objectionToAdd}(A,?Y).\\
&\text{allCandi}\text{dates}(A,?Y)\text{:- allCoord}(A,?Z),\\
&\hspace{1cm}\text{agreeToAdd}(?Z,?Y).\\
&\text{objectionToAdd}(A,?Y)\text{:- allCoord}(A,?Z),\\
&\hspace{1cm}\text{disagreeToAdd}(?Z,?Y).\\
&\text{disagreeToAdd}(A,?Y)\text{:- allCandidates}(A,?Y),\\
&\hspace{1cm}\neg \text{agreeToAdd}(A,?Y).\\
&\text{allCoord}(A,?Y)\text{:- allCoord}(A,?Z),\\
&\hspace{1cm}\text{coord}(?Z,?Y).\\
&\text{allCoord}(A,A).\\
&\text{coord}(A,B).\\
&\text{coord}(B,C).\\
&\text{coord}(C,B).\\
&\text{coord}(C,A).\\
&\text{agreeToAdd}(A,D).\\
&\text{disagreeToAdd}(A,E).\\
&\text{disagreeToAdd}(B,F).\\
&\text{disagreeToAdd}(C,F).\\
\end{split}
\end{equation*}
}\\ \hline
\end{tabular}
\end{center}
\end{table}

\begin{center}
\begin{table}[ht]
\footnotesize
\begin{center}
\caption{Execution time of the program on the XSB, SMODELS, and DLV 
systems}
\begin{tabular}{|c|c|c|c|c|c|c|c|c|c|}
\hline
 & \multicolumn{3}{c|}{10 Coords}& \multicolumn{3}{c|}{30 Coords}& 
\multicolumn{3}{c|}{50 Coords}\\
\cline{2-10}
& \multicolumn{3}{c|}{Num. of Iterations}& \multicolumn{3}{c|}{Num. of Iterations}& 
\multicolumn{3}{c|}{Num. of Iterations}\\
\cline{2-10}
 & 1 & 10 & 20 & 1 & 10 & 20 & 1 & 10 & 20 \\
\hline
\hline
DLV     & 0.05s & 0.81s & 1.54s & 0.06s & 0.83s & 1.55s & 0.07s & 0.86s 
& 1.60s \\
SMODELS & 0.12s & 1.22s & 2.32s & 0.16s & 1.35s & 2.66s & 0.19s & 1.53s 
& 2.94s \\
\cline{2-10}
XSB     & \multicolumn{9}{c|}{$\approx$ 0} \\
\hline
\end{tabular}
\label{result}
\end{center}
\end{table}
\end{center}

\end{document}